\documentclass[prl,reprint,twocolumn,superscriptaddress,showpacs,floatfix]{revtex4-1}

\usepackage{amsmath}
\usepackage{mathrsfs}
\usepackage{txfonts}
\usepackage{amssymb}
\usepackage{graphicx}
\usepackage{hyperref}
\usepackage{ulem}
\usepackage{overpic}
\usepackage{psfrag}
\usepackage{tabularx}
\usepackage{array}
\usepackage{placeins}
\newcommand{\PreserveBackslash}[1]{\let\temp=\\#1\let\\=\temp}

\usepackage{color}

\makeatletter

\begin{document}
	
	\title{Instability of the Luttinger liquids towards an exotic quantum state of matter with highly degenerate ground states: an anisotropic extension of the ferromagnetic spin-1 biquadratic model}
	
	\author{Qian-Qian Shi}
	\affiliation{Centre for Modern Physics, Chongqing University, Chongqing 400044, The People's Republic of China}
	\author{Yan-Wei Dai}
	\affiliation{Centre for Modern Physics, Chongqing University, Chongqing 400044, The People's Republic of China}
	\author{Sheng-Hao Li}
	\affiliation{Centre for Modern Physics, Chongqing University, Chongqing 400044, The People's Republic of China}
	\author{Huan-Qiang Zhou}
	\affiliation{Centre for Modern Physics, Chongqing University, Chongqing 400044, The People's Republic of China}
	
	\begin{abstract}
		An extensive investigation, both numerical and analytical, is performed for an anisotropic extension of the ferromagnetic spin-1 biquadratic model. The ground state phase diagram accommodates three symmetry-protected trivial phases, three coexisting fractal phases and six Luttinger liquid phases. A novel universality class arises from an instability of a Luttinger liquid towards an exotic quantum state of matter with infinitely degenerate ground states.  The latter in turn is a scale-invariant quantum state of matter, which may be attributed to the coexistence of ${\rm SU}(2)$ spontaneous symmetry breaking with one type-B Goldstone mode on the characteristic line: $J_y=J_z$, and ${\rm U}(1)$ spontaneous symmetry breaking without any gapless Goldstone mode on the characteristic line $J_x/J_z=0$, together with their cyclic permutations with respect to $x$, $y$ and $z$.
	\end{abstract}
	\maketitle
	
	{\it Introduction.}- In  the conventional Landau-Ginzburg-Wilson paradigm, spontaneous symmetry breaking (SSB)~\cite{anderson} is a fundamental notion, which results in different types of long-range order to characterize distinct quantum states of matter.  In this regard, a fundamental theorem {\it a la} Coleman~\cite{coleman}, which in turn is a quantum counterpart of
	the Mermin-Wagner theorem in two-dimensional classical statistical systems, states that no continuous symmetry is spontaneously broken in one-dimensional quantum many-body systems, simply due to strong quantum fluctuations.
	As a consequence, the Kosterlitz-Thouless (KT) transitions~\cite{kt}, as a prototypical topological phase transition, must be beyond any SSB description.  In fact, the KT transitions describe an instability of the Luttinger liquid phases (LLs), due to marginal perturbations, towards a gapful phase, with or without $Z_2$ SSB order.
	Another type of instabilities of the LL phases is the Pokrovsky-Talapov (PT) phase transitions~\cite{pt}, which essentially originate from the energy-level crossings, with a remarkable feature being the absence of conformal invariance at a transition point.
	
	A natural question arises as to whether or not the KT and PT transitions exhaust all possible instabilities of the LL phases towards distinct quantum states of matter under different types of perturbations.  As evidenced by the Bethe ansatz exact solutions for quantum spin-1/2 XXZ chain in an external magnetic field~\cite{xxz,xxz2,xxz3,xxz4}, there might be some other types of instabilities of the LL phases~\cite{fm}.   Actually, two remarks are in order.  First, in addition to both the KT and PT transitions, this model exhibits a type of quantum phase transitions (QPTs) interpolating between the KT and PT transitions.  Second, a QPT occurs at the $\rm{SU(2)}$ ferromagnetic point, which features highly degenerate and highly entangled ground states~\cite{popkov}.
	
	We aim to address this intriguing issue through an extensive investigation into an anisotropic extension of the ferromagnetic spin-1 biquadratic model, both numerical and analytical. Numerical simulations are carried out in terms of infinite time evolving block decimation (iTEBD)~\cite{vidal}, which is based on infinite matrix product state (iMPS) representation.  Our results suggest that an exotic quantum state of matter with infinitely degenerate ground states arises, which may be attributed to the coexistence of ${\rm SU}(2)$ SSB with one type-B Goldstone mode (GM)~\cite{watanabe0,watanabe} on the characteristic line $J_y=J_z$, and ${\rm U}(1)$ SSB without any gapless GM~\cite{U1SSB}  on the characteristic line $J_x/J_z=0$,  together with their cyclic permutations with respect to $x$, $y$ and $z$.  Both SSB patterns survive quantum fluctuations in one spatial dimension, in contrast to SSB with type-A GMs subject to the Mermin-Wagner-Coleman theorem~\cite{coleman}.
	A peculiar feature of the model is that it possesses distinct symmetries with varying anisotropic coupling parameters, up to a $\rm{SU(3)}$ symmetry at the isotropic point.
	The ground state phase diagram accommodates three symmetry-protected trivial (SPt) phases, three coexisting fractal (CF) phases - scale-invariant quantum states of matter with infinitely degenerate ground states, and six LL phases.  In addition, QPTs between the LL phases and the SPt phases are identified to be in the KT universality class.
	
	{\it  An anisotropic extension of the ferromagnetic spin-1 biquadratic model.} - The model Hamiltonian takes the form
	\begin{equation}
		H(J_x,J_y,J_z)=\sum_{j}{(J_xS_j^x S_{j+1}^x+J_yS_j^yS_{j+1}^y+J_zS_j^zS_{j+1}^z)^2}. \label{xyz2}
	\end{equation}
	Here, $S_j^x$, $S_j^y$, and $S_j^z$  are the spin-$1$ operators at a lattice site $j$, and $J_x$, $J_y$, and $J_z$ denote the coupling parameters describing anisotropic interactions.
	The model is symmetric under a unitary transformation: $S^x_j\rightarrow(-1)^jS^x_j$, $S^y_j\rightarrow(-1)^jS^y_j$, $S^z_j\rightarrow S^z_j$, accompanied by $J_x\rightarrow J_x$, $J_y\rightarrow J_y$ and $J_z\rightarrow -J_z$, or its counterparts under a cyclic permutation with respect to $x, y$ and $z$.
	Therefore, we may restrict our discussion to the parameter region: both $J_x/J_z$ and $J_y/J_z$ are non-negative. It enjoys distinct symmetry groups with varying coupling parameters.
	Specifically,  a symmetry group ${\rm U}(1) \times {\rm U}(1)$, for the region
	$0 \leq J_x\leq J_z$, $0 \leq J_y\leq J_z$, and $J_x\leq J_y$, is generated by $K_{yz}$ and $K_{x}$, with $K_{yz}= \sum _j (-1)^{j+1} [(S_j^{y})^2-(S_j^{z})^2]$ and $K_{x}=\sum _j (-1)^{j+1} (S_j^{x})^2$, respectively.
	On the characteristic line  $J_y=J_z$, a $\rm{SU(2)}$ symmetry group is generated from
	$\Sigma_x=\sum_jS_j^x/2$,  $\Sigma_y=K_{yz}/2$ and $ \Sigma_z=\sum_j(-1)^{j+1}(S_j^yS_j^z+S_j^zS_j^y)/2$, satisfying $[\Sigma_{\lambda}, \Sigma_{\mu}] = i \varepsilon_{\lambda \mu \nu} \Sigma_{\nu}$, where $\varepsilon _{\lambda \mu \nu}$ is a completely antisymmetric tensor, with $\varepsilon_{xyz}=1$, and $\lambda, \mu, \nu = x,y,z$, and a ${\rm U(1)}$ symmetry group is generated by $K_x=\sum _j (-1)^{j+1}(S_j^x)^2$.
	Then, the generators of the symmetry group ${\rm U}(1) \times {\rm U}(1)$ in the other regions, and the $\rm{SU(2)}\times {\rm U(1)}$ symmetry group on the characteristic lines $J_x=J_y$ and $J_z=J_x$, follow from the duality transformations induced from the symmetric group ${\rm S}_3$,  arising from cyclic permutations with respect to $x$, $y$, and $z$.
	As a result, a $\rm{SU(3)}$ symmetry group emerges at the isotropic point $J_x = J_y = J_z$.
	Hence, the entire parameter region is partitioned into six different regimes, which are dual to each other.
	More details about the symmetry groups and dualities are described in Sec. A and Sec. B of the Supplemental Material (SM), respectively.

\begin{figure}
		\includegraphics[angle=0,totalheight=5cm]{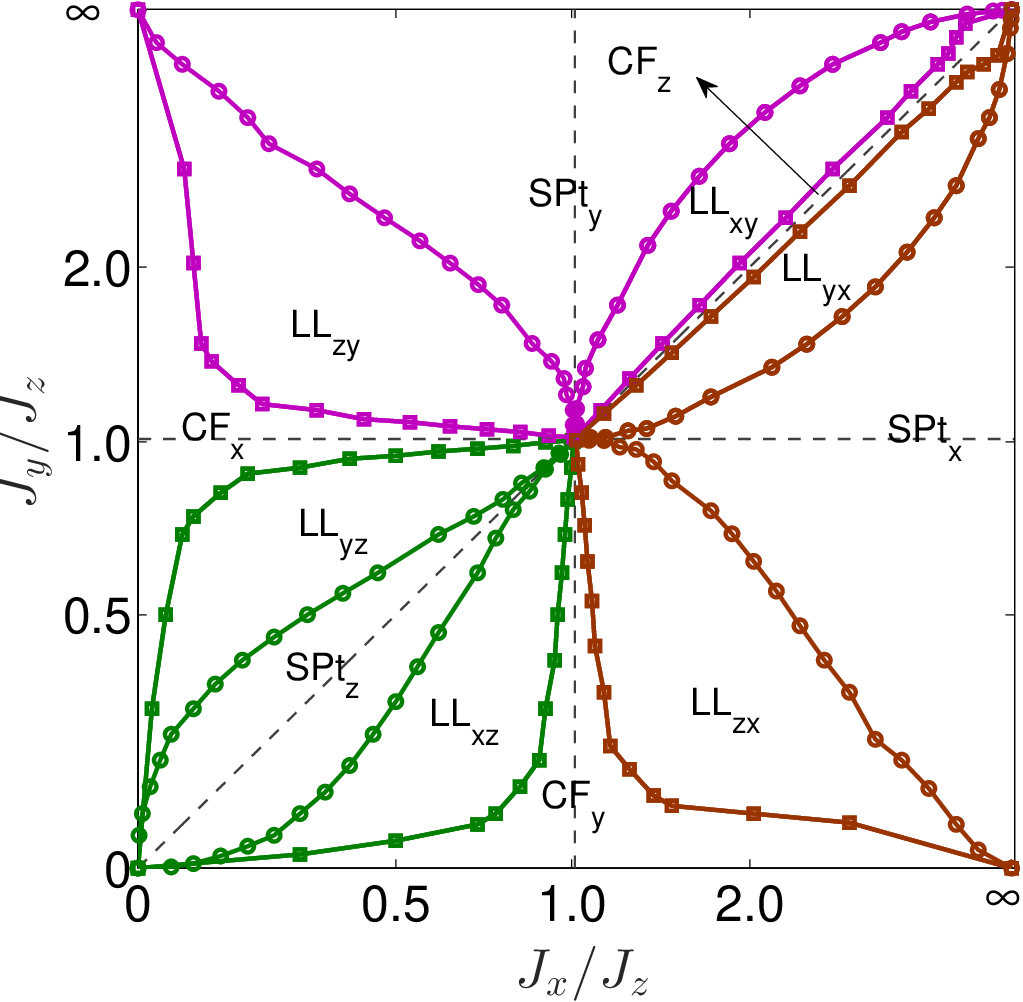}
		\caption{ Ground state phase diagram for an anisotropic extension of the ferromagnetic spin-1 biquadratic model. We restrict our attention to $J_x/J_z \geq 0$ and $J_y/J_z \geq 0$, due to a  symmetric consideration. Here, a solid line indicates a phase transition line.
			The model accommodates twelve distinct phases: three CF phases labeled as $\rm{CF_{x}}$, $\rm{CF_{y}}$ and $\rm{CF_{z}}$, six LL phases labeled as $\rm{LL_{xy}}$, $\rm{LL_{yz}}$, $\rm{LL_{zx}}$, $\rm{LL_{yx}}$, $\rm{LL_{xz}}$ and $\rm{LL_{zy}}$,  and three SPt phases labeled as $\rm{SPt_{x}}$, $\rm{SPt_{y}}$  and $\rm{SPt_{z}}$, respectively.
			Note that both horizontal and vertical axes are in a scale, defined by $\arctan\;(J_x/J_z)$ and $\arctan\;(J_y/J_z)$, respectively.
		}\label{phasediagram}
	\end{figure}

	{\it Ground state phase diagram.} - The ground state phase diagram is plotted in Fig.~\ref{phasediagram}, which
	accommodates twelve distinct phases: three CF phases labeled as $\rm{CF_{x}}$, $\rm{CF_{y}}$ and $\rm{CF_{z}}$, six LL phases labeled as $\rm{LL_{xy}}$, $\rm{LL_{yz}}$, $\rm{LL_{zx}}$, $\rm{LL_{yx}}$, $\rm{LL_{xz}}$ and $\rm{LL_{zy}}$,  and three SPt phases labeled as $\rm{SPt_{x}}$, $\rm{SPt_{y}}$  and $\rm{SPt_{z}}$, respectively.  As it turns out, an exotic type of QPTs arises from an instability of a LL phase towards a CF phase.  In addition, QPTs between the LL phases and the SPt phases are identified to be in the KT universality class.
	
	The strategy we adopt to map out the entire phase diagram is to choose one of the six dual regimes as a principal regime, preferably a regime with finite extents, and to employ powerful tensor network algorithms based on the iMPS representations to simulate the model in this regime. Throughout this work, the principal regime is chosen to be $0 \leq J_x\leq J_z$, $0 \leq J_y\leq J_z$, and $J_x\leq J_y$.  For our purpose,
	the iTEBD algorithm~\cite{vidal} is exploited to locate phase boundaries in the principal regime.  This is achieved in terms of the (block) entanglement entropy for all the twelve phases, the local order parameters for the CF phases, the pseudo local order parameters for the LL phases and the non-local order parameters for the SPt phases.
	Once distinct phases are detected in the chosen principal regime, the entire ground state phase diagram follows from the duality transformations.
	
	{\it Characterization of distinct phases.} - Now we turn to a full characterization of
	the $\rm{CF}$ phases, the LL phases and the SPt phases.
	
	(a) A CF phase as an exotic quantum state of matter with highly degenerate ground states:  From the perspective of  the duality transformations and the symmetry groups, $J_x=0$ ($J_y=0$) and $J_y=J_z$ ($J_z=J_x$) are two characteristic lines.  Actually, the model features highly degenerate ground states on the two characteristic lines.
	
	On the characteristic line $J_x=0$ with $J_y/J_z>0$, a brute force calculation shows that a factorized ground state, with the ground state energy per site $e$ being equal to $0$,  takes the form
	\begin{equation}
		|\Psi_f\rangle=\bigotimes_m |v_1v_2\rangle _m,
		\label{v12f}
	\end{equation}
	where $|v_1v_2\rangle _m = |v_1\rangle_{2m-1} |v_2\rangle _{2m}$,
	with $|v_1 \rangle_{2m-1}$ and  $|v_2 \rangle_{2m}$ being a vector in a local spin space at lattice sites $2m-1$ and $2m$, respectively,
	\begin{eqnarray}
		|v_1\rangle_{2m-1}=&p|0_y\rangle_{2m-1} + q |0_z\rangle_{2m-1},\nonumber \\
		|v_2\rangle_{2m}= &s|0_y\rangle_{2m} +t|0_z\rangle_{2m}.
		\label{v12}
	\end{eqnarray}
	Here, $q\!=\!\exp{(i\delta)}\sqrt{1-p^2}$, $\!s\!=\!\sqrt{1-p^2}J_y\;/\sqrt{(1-p^2)\!J_y^2\!+\!p^2\!J_z^2\!}$, and $t=\exp{(-i\delta)}pJ_z\;/\sqrt{(1-p^2)J_y^2+p^2J_z^2}$, with $p$ and $\delta$ being two (real) free parameters, and $|0_y\rangle_{2m-1/2m}$ and $|0_z\rangle_{2m-1/2m}$ are eigenvectors, with an eigenvalue being zero, for the spin operators $S^y_{2m-1/2m}$ and $S^z_{2m-1/2m}$, respectively.
	Generally, $|\Psi_f\rangle$ is not translation-invariant, indicating that the one-site translational symmetry is spontaneously broken, except for a special case:  $p=\sqrt{J_y/(J_y+J_z)}$ and $\delta=0$.
	The latter, denoted as $|{\rm TIGS}\rangle$, becomes a factorized ground state~(\ref{v12f}), with $p=s=q=t=\sqrt{2}/2$, if $J_y=J_z$.
	That is, the model admits a two-parameter family of degenerate factorized ground states on the characteristic line $J_x=0$. This allows a group representation-theoretic interpretation
	(for more details, cf. Sec.\;C of the SM). In fact, the degenerate factorized ground states (\ref{v12f}) simply follow from the action of the ${\rm U}(1)$ symmetry group element $V=\exp{(\alpha K_{yz})}$  on the
	translation-invariant state $|{\rm TIGS}\rangle$.   Note that $\alpha$ is a complex number, meaning that both the unitary and non-unitary realizations of the symmetry group ${\rm U}(1)$, generated by $K_{yz}$, are allowed.
	Given that the degenerate ground states $|\Psi_f\rangle$ for different values of $p$ and $\delta$ are asymptotically orthogonal to each other, the ${\rm U}(1)$ symmetry, generated by $K_{yz}$, is spontaneously broken in the thermodynamic limit~\cite{U1SSB},  thus constituting a counter-example to the theorem in Ref.~\cite{schafer}.

	A remarkable fact is that, in the thermodynamic limit, the factorized ground states $|\Psi_f\rangle$,
	constructed for the Hamiltonian (\ref{xyz2}) on the characteristic line $J_x=0$, also constitute ground states for the Hamiltonian (\ref{xyz2}) in the entire $\rm{CF_{x}}$ phase, with the ground state energy per site being $J_x^2$.
	That is, $H|\Psi_f\rangle= L J_x^2|\Psi_f\rangle$ is valid in the entire phase, if $L\rightarrow \infty$.
	Indeed, for a finite-size system with the size being $L$, the ground state energy per site $e(L)$ in the $\rm{CF_x}$ phase takes the form 
	\begin{equation}
		e(L)= J_x^2-A\frac{e^{\eta/ L}}{L}-Be^{-\kappa L}, \label{fsc}
	\end{equation}
	where $A$, $B$, $\eta$ and $\kappa$ are real and positive.
	We stress that the finite-size corrections to the ground state energy per site may be justified from the fact that there are a sequence of quantum states  $|\psi_k\rangle$ ($k=0,1,\cdots,q-1$)~\cite{U1SSB}, with $q=L+1$, satisfying the H-orthogonality~\cite{szb}. More precisely, the $q$ H-orthogonal states  $|\psi_k\rangle$ ($k=0,1,\cdots,q-1$), defined as $|\psi_k\rangle \equiv (V_q)^k |{\rm TIGS}\rangle$, where $V_q$ denotes an operator $V_q=\prod _j\exp(i2\pi K_{yz}^j/q)$. Indeed, $V_q$ itself is an element of a cyclic group ${\rm Z}_{q}$, which turns out to be a subgroup of the symmetry group $\rm{U}(1)$ generated by $K_{yz}$. As argued in Ref.~\cite{U1SSB}, two length scales are  competing with each other in the ${\rm CF}_x$ phase.  One is involved in the second term originating from the emergent permutation symmetry in the ground state subspace,
	which in turn is relevant to a gapped GM when the symmetry group $\rm{SU}(2)\times \rm{U}(1)$ on the
	characteristic line $J_y = J_z$ is explicitly broken to $\rm{U}(1)\times \rm{U}(1)$ in the  ${\rm CF}_x$ phase, away from the
	characteristic line $J_y = J_z$. The other is involved in the third term, originating from an alternative SSB pattern for $\rm{U(1)}$. 
	Indeed, in the thermodynamic limit, the symmetry group ${\rm  U(1)}$, viewed as a limit of $Z_q$ when $q \rightarrow \infty$, is spontaneously broken, thus leading to an alternative SSB pattern for $\rm{U(1)}$, with no emergent gapless GM, in sharp contrast to the Goldstone theorem~\cite{goldstone}. 
	Our numerical simulations in terms of the finite-size density matrix renormalization group (DMRG) algorithm~\cite{dmrg,dmrg2} confirm the finite-size corrections (for more details, cf. Sec.\;D of the SM).

	The iTEBD simulations of the model (\ref{xyz2}) in the $\rm{CF_x}$ phase, with $J_y/J_z=0.7$ as an example, lend further support.
	As discussed in  Sec. D of the SM, both the entanglement entropy $S(\chi)$ and the amplitude of the local order parameter, denoted as $|\langle Q_j \rangle|$, exhibit randomness arising from highly degenerate ground states in the $\rm{CF_x}$ phase, when the iTEBD algorithm is implemented. Here, $Q_j$ is defined as $Q_{2m-1}=S_{2m-1}^yS_{2m-1}^z$ for odd sites $j=2m-1$ and $Q_{2m}=-S_{2m}^zS_{2m}^y$ for even sites $j=2m$.
	Note that the real and imaginary parts of  $\langle Q_j \rangle$ are $ \langle (-1)^{j+1}(S^y_jS^z_j+S^z_jS^y_j)/2 \rangle$ and $ \langle S^x_j/2 \rangle$, respectively. That is, for the real part, the local operator is alternating between odd and even sites, whereas for the imaginary part, the local operator is uniform. Further, the randomness originates from the fact that both the unitary and non-unitary realizations of the symmetry group ${\rm U}(1)$, generated by $K_{yz}$, are spontaneously broken in the $\rm{CF_x}$ phase.
	This may be recognized as a smoking-gun signature for the $\rm{CF_x}$ phase.
	
		\begin{figure}[htbp]
		\includegraphics[angle=0,width=0.3\textwidth]{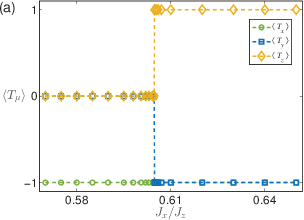}\\
		\includegraphics[angle=0,width=0.3\textwidth]{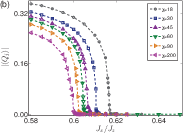}\\
		\caption{ For $J_y/J_z=0.7$, (a) the non-local order parameters $\langle T_{\mu}\rangle$ as a function of $J_x/J_z$,  with the bond dimension $\chi=60$ and  (b) the amplitude of the pseudo local order parameter, denoted as $|\langle Q_j \rangle|$, as a function of $J_x/J_z$, with the bond dimension $\chi=18$, $30$, $45$, $60$, $90$ and $200$.
			The non-local order parameters $\langle T_{\mu}\rangle$ take different values in different phases:  $(-1,\;-1,\;1)$ in the $\rm{SPt_{z}}$ phase and $(-1,\;0,\;0)$ in the $\rm{LL_{yz}}$ phase, and the amplitude of the pseudo local order parameter, denoted as $|\langle Q_j \rangle|$, takes nonzero values in the $\rm{LL_{yz}}$ phase, and is zero in the $\rm{SPt_{z}}$ phase.
			Therefore, $\langle T_{\mu}\rangle$ and  $\langle Q_j \rangle$ may be used to distinguish the $\rm{SPt_{z}}$ phase from the $\rm{LL_{yz}}$ phase.
		}\label{order}
	\end{figure}

	We emphasize that SSB from ${\rm U(1)} \times {\rm U(1)}$ to ${\rm U(1)}$ in the $\rm{CF_x}$ phase smoothly evolves into SSB from $ {\rm SU}(2)\times  {\rm U}(1)$ to  ${\rm U}(1) \times{\rm U}(1)$~\cite{FDGM} on the characteristic line $J_y=J_z$.  Meanwhile, $|{\rm TIGS}\rangle$ becomes $\bigotimes_j|1_x\rangle$.
	The latter is the highest weight state for the symmetry group  ${\rm SU}(2)$.
	As a result, SSB with one type-B GM~\cite{watanabe} arises.  In other words, ${\rm SU}(2)$ SSB with one type-B GM on the characteristic line $J_y=J_z$ coexists with ${\rm U}(1)$ SSB without any gapless GM on the characteristic line $J_x=0$, given both the characteristic lines are in the {\it same} phase.  This yields highly degenerate and highly entangled ground states on the characteristic line $J_y=J_z$, with the fractal dimension $d_f$\cite{doyon} being identical to the number of type-B GMs: $d_f=N_B$ and $N_B=1$.
	We remark that SSB from ${\rm U(1)} \times {\rm U(1)}$ to ${\rm U(1)}$ also occurs in the ${\rm CF_y}$ and ${\rm CF_z}$ phases, together with ${\rm SU}(2)$ SSB with one type-B GM on the characteristic lines $J_z=J_x$ and $J_x=J_y$, as follow from the duality transformations.
	At the isotropic point ($J_x=J_y=J_z$), SSB from the staggered ${\rm SU(3)}$ to ${\rm U(1)}\times {\rm U(1)}$ occurs, with the fractal dimension $d_f=2$, since the number of type-B GMs is equal to 2~\cite{staggeredsu3}.
	
	(b) The SPt phases: For $J_y/J_z=0.7$, with $J_x/J_z>J_x^c/J_z$ ($J_x^c\sim0.6$), there exists a SPt phase, denoted as the $\rm{SPt_{z}}$ phase, as seen in  Fig.~\ref{phasediagram}.
	To characterize the $\rm{SPt_{z}}$ phase, the non-local order parameters $\langle T_{\mu} \rangle$ are introduced.  Following Ref.~\cite{pollmann2}, $\langle T_{\mu} \rangle$ are defined through the combined operation of the site-centered inversion $I$ with the $\pi$-rotation $R_{\mu}=\exp{(i\pi S_\mu)}$ around the $\mu$-axis, with $\mu=x,y$, and $z$, in the spin space: $T_{\mu}=I \cdot R_{\mu}$.
	A detailed procedure to evaluate the non-local order parameters $\langle T_{\mu} \rangle$ from the iMPS representation is described in  Sec. E of the SM.
	In Fig.~\ref{order}, $\langle T_{\mu}\rangle$ is plotted as a function of $J_x/J_z$ for fixed $J_y/J_z=0.7$, which succeeds in distinguishing  the $\rm{LL_{yz}}$ phase from the $\rm{SPt_{z}}$ phase.
	Indeed, $(\langle T_x\rangle,\langle T_y\rangle,\langle T_z\rangle)$ take different values in different phases:  $(-1,\;-1,\; 1)$ in the $\rm{SPt_{z}}$ phase and $(-1,\;0,\;0)$ in the $\rm{LL_{yz}}$ phase.   As a consistency check, the phase transition points, thus detected for each value of the bond dimension $\chi$, match that from the entanglement entropy $S(\chi)$.
	
	Since the duality transformations are induced from the symmetric group $S_3$, three SPt phases $\rm{SPt_{x}}$, $\rm{SPt_{y}}$ and $\rm{SPt_{z}}$ are distinguished from each other, with the non-local order parameters $\langle T_{\mu} \rangle$ taking different values: $(1,\; -1,\; -1)$ for $\rm{SPt_{x}}$, $(-1,\; 1,\; -1)$ for $\rm{SPt_{y}}$, and $(-1,\; -1,\; 1)$ for $\rm{SPt_{z}}$, respectively.

	(c) The LL phases: According to the Mermin-Wagner-Coleman theorem, no local order parameter exists to characterize any LL phase, in the sense that no continuous SSB order survives quantum fluctuations in one spatial dimension. However, in the iTEBD simulations, only accessible is a finite value of the bond dimension $\chi$. This amounts to suppressing quantum fluctuations, which in turn yields pseudo continuous SSB. As a result, we may introduce a pseudo local order parameter to characterize a LL phase ~\cite{wang,dai0} (see also Sec.\;F of the SM for the details).
	
	In the ${\rm LL}_{yz}$ phase, pseudo SSB emerges for a unitary realization of the symmetry group ${\rm U}(1)$,
	generated by $K_{yz}$, which results in the pseudo local order parameter.
	The pseudo local order parameter  $\langle Q_j \rangle$ in the $\rm{LL_{yz}}$ phase is chosen to be identical to that in the ${\rm CF}_{x}$.
	Note that a unitary transformation $W=\exp(i\varepsilon K_{yz})$ induces a phase factor in the pseudo local order parameter $\langle Q_j \rangle$: $\exp(i\;2\varepsilon)$.
	In Fig.~\ref{order}~(b), we plot the amplitude of the pseudo local order parameter, denoted as $|\langle Q_j \rangle|$, for $J_y/J_z=0.7$, with the bond dimension $\chi=18$, $30$, $45$, $60$, $90$ and $200$, respectively.  We emphasize that the phase transition points, detected from the pseudo local order parameter $\langle Q_j \rangle$ for different values of the bond dimension $\chi$, are identical to that determined from the entanglement entropy $S(\chi)$.
	
\begin{figure}
		\includegraphics[angle=0,width=0.3\textwidth]{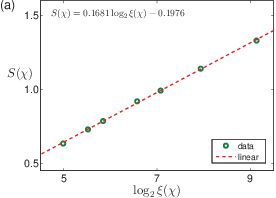}
		\vspace{1cm}
		\includegraphics[angle=0,width=0.3\textwidth]{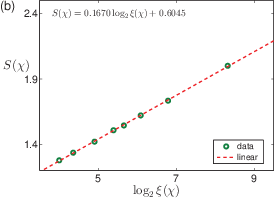}
		\caption{ Scaling of the entanglement entropy $S(\chi)$ with the logarithm of the correlation length $\log_2\xi(\chi)$  (a) for $J_x/J_z=0.5$ and $J_y/J_z=0.7$, which is located in the $\rm{LL_{yz}}$ phase and (b) at the pseudo phase transition points for fixed $J_y/J_z=0.7$, which are the peak positions of entanglement entropy $S(\chi)$ for each value of the bond dimension $\chi$. Here, the bond dimension  $\chi$ ranges from $18$ to $200$. We extract central charge $c$ to be (a) $c=1.009$ and (b) $c=1.002$, respectively.
		}\label{centralc}
	\end{figure}

	As a further confirmation, we  randomly choose different values of the coupling parameters, which are located in this phase, to extract central charge $c$ from a finite-entanglement scaling analysis~\cite{centralchargescaling}.  According to a prediction from conformal field theory, the entanglement entropy $S(\chi)$ scales as
	\begin{equation}
		S(\chi)=\frac{c}{6}\log_2\xi(\chi)+a.
		\label{finites}
	\end{equation}
	Here, the correlation length  $\xi(\chi)$ scales with the bond dimension $\chi$ as $\xi(\chi)=b\chi^{\kappa}$, with $\kappa$ being an exponent to be determined, and $a$ and $b$ being some constants.  As an illustration, we consider $J_x/J_z=0.5$ and $J_y/J_z=0.7$.  Our simulations yield central charge $c=1.009$, with the bond dimension $\chi$ ranging from $18$ to $200$, as shown in Fig.~\ref{centralc} (a). More numerical results are listed in Table~\ref{tab1}, for five randomly chosen values of the parameters $J_x/J_z$ and $J_y/J_z$. Note that, the relative errors, as a measure of the deviation from the exact value $c=1$, are less than 2 percent.  The consistent results for central charge $c$ also follow from a finite-size approach (cf. Sec.\;G of the SM).
	
	\begin{table}[htbp]
		\centering
		\caption{Central charge $c$, extracted from the iTEBD simulations, for five chosen points in the $\rm{LL_{yz}}$ phase. }
		\vspace{3mm}
		\label{tab1}
		\begin{tabular}{|c|c|c|c|c|c|}
			\hline
			($J_x/J_z$,$J_y/J_z$)&(0.15,0.5)&(0.3,0.6)&(0.2,0.7)&(0.5,0.7)&(0.7,0.8)\\
			\hline
			$c$&0.985&1.017&0.983&1.009&0.997 \\
			\hline
		\end{tabular}
	\end{table}

\begin{figure}[htbp]
		\includegraphics[angle=0,width=0.31\textwidth]{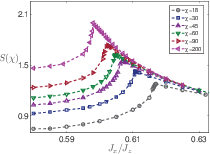}
		\caption{For fixed $J_y/J_z=0.7$, the entanglement entropy $S(\chi)$ as a function of $J_x/J_z$ for the bond dimension $\chi$ ranging from 18 to 200. The phase transitions from the $\rm{LL_{yz}}$ phase to the $\rm{SPt_z}$ phase are detected from the presence of the peaks in the entanglement entropy $S(\chi)$.  Note that the peak positions shift when the bond dimension $\chi$ increases.
		}\label{entropy}
	\end{figure}

	{\it The KT transitions from the LL to the symmetry-protected trivial phases.} - Here, we focus on a phase transition between the $\rm{LL_{yz}}$ phase and the $\rm{SPt_z}$ phase for $J_y/J_z=0.7$.  As shown in Fig.~\ref{entropy}, the entanglement entropy $S(\chi)$ as a function of $J_x/J_z$ for  $J_y/J_z=0.7$,  exhibits a peak for each value of the bond dimension $\chi=18$, $30$, $45$, $60$, $90$ and $200$, respectively.  This indicates that a continuous QPT occurs, as follows from the continuity of the entanglement entropy $S(\chi)$ at a peak position $J_x^c/J_z$.  The peak positions are supposed to converge to a critical point. Our task is to determine what universality class it belongs to. To accomplish this, we perform a finite-entanglement scaling analysis (\ref{finites}) to extract central charge $c$ at a critical point~\cite{centralchargescaling}.

	Our results from the iTEBD simulations are plotted in Fig.~\ref{centralc} (b), with the bond dimension $\chi$ ranging from $18$ to $200$, respectively.  The peak values of the entanglement entropy $S(\chi)$ scale with the correlation length $\xi(\chi)$ as $S(\chi)=0.1670\log_2\xi(\chi)+0.6045$, which yields $c=1.002$.  Hence, we are led to conclude that the phase transitions between the $\rm{LL_{yz}}$ phases and the $\rm{SPt_{z}}$ phases are the KT transitions.

	{\it Instability of the Luttinger liquids towards the CF phases.} - In addition to the KT transitions from the LL phases to the SPt phases, there is another type of QPTs emerging from the instability of the LL phases towards the CF phases.  Here, it is proper to stress that simulating the Hamiltonian (\ref{xyz2}) in the CF phases in terms of the iTEBD algorithm is really challenging, although it is tractable analytically. A remarkable fact is that the entanglement entropy $S(\chi)$ exhibits different behaviors for two distinct phases: it takes a definitive value in the LL phases, and a random value in the CF phases, respectively.  In fact, this observation may be used to signal QPTs from the LL phases to the CF phases.  Randomness also occurs in $|\langle Q_j\rangle|$, which is chosen to be a local pseudo order parameter in the $\rm{LL_{yz}}$ phase, and a local order parameter in the $\rm{CF_x}$ phase.
	For fixed $J_y/J_z=0.7$,  $|\langle Q_j\rangle|$ exhibits distinct behaviors in the $\rm{LL_{yz}}$ phase and the $\rm{CF_x}$ phase: in the $\rm{LL_{yz}}$ phase,  $|\langle Q_j\rangle|$ takes a definite value and varies smoothly with $J_x/J_z$, whereas in the $\rm{CF_x}$ phase, $|\langle Q_j\rangle|$ varies randomly with $J_x/J_z$ (cf. Sec. D of the SM).
	A physical interpretation for the distinct behaviors is as follows. In the ${\rm LL}_{yz}$ phase, only pseudo SSB for a unitary realization of the symmetry group ${\rm U}(1)$, generated by $K_{yz}$, occurs in the iTEBD simulations, which tends to vanish when the bond dimension $\chi$ increases. In contrast, an alternative SSB pattern for $\rm{U(1)}$, with no accompanied gapless GM, emerges, thus leading to infinitely degenerate ground states in the $\rm{CF_x}$ phase. The same argument works for the other  $\rm{CF}$ phases as a result of the duality transformations. As a consequence, the instability of a LL phase towards a CF phase, which is scale-invariant but not conformally invariant, leads to a novel universality class, different from the KT and PT transitions.

{\it Acknowledgement.} - We acknowledge enlightening discussions with Murray Batchelor, Sam Young Cho, John Fjaerestad, Javier Rodriguez Laguna, Ian McCulloch, and German Sierra.

\newpage
\section*{Supplementary Material}
\twocolumngrid
\setcounter{page}{1}
\setcounter{equation}{0}
\setcounter{figure}{0}
\setcounter{table}{0}
\renewcommand{\theequation}{S\arabic{equation}}
\renewcommand{\thefigure}{S\arabic{figure}}
\renewcommand{\thetable}{S\arabic{table}}
\renewcommand{\bibnumfmt}[1]{[S#1]}
\renewcommand{\citenumfont}[1]{S#1}

\subsection{Symmetry groups }\label{symmetrygroup}

The model Hamiltonian (1) is peculiar, in the sense that it enjoys distinct symmetry groups when the anisotropic couplings $J_x$, $J_y$ and $J_z$ are varied.

In the entire parameter space, there are ${\rm U}(1) \times {\rm U}(1)$ symmetry groups, constructed from the three ${\rm U(1)}$ symmetry groups ${\rm U(1)_{xy}}$, ${\rm U(1)_{yz}}$ and ${\rm U(1)_{zx}}$, which are generated by $K_{xy}$, $K_{yz}$ and $ K_{zx}$,
with $K_{xy}= \sum _j (-1)^{j+1} [(S_j^{x})^2-(S_j^{y})^2]$, $K_{yz}=\sum _j (-1)^{j+1} [(S_j^{y})^2-(S_j^{z})^2]$ and $K_{zx}=\sum _j (-1)^{j+1} [(S_j^{z})^2-(S_j^{x})^2]$, respectively.
In fact, only two among the three ${\rm U(1)}$ symmetry groups are independent to each other, due to the fact that $K_{xy}+K_{yz}+K_{zx}=0$.
Specifically, in the region with $0 \leq J_x\leq J_z$, $0 \leq J_y\leq J_z$, and $J_x\leq J_y$,  the symmetry group ${\rm U}(1) \times {\rm U}(1)$ is generated by $K_{yz}$ and $K_{x}$,  with $K_{x}=\sum _j (-1)^{j+1} (S_j^{x})^2$.
In addition, the symmetry groups in the other regions follow from the duality transformations, induced from the symmetric group $S_3$, arising from the cyclic permutations with respect to $x$, $y$, and $z$.

On each of the three characteristic lines: (1) $J_x=J_y$, (2) $J_y=J_z$ and (3) $J_z=J_x$, a ${\rm SU(2)}$ symmetry group generated by $\Sigma_x$, $\Sigma_y$, and $\Sigma_z$ emerges, satisfying
$[\Sigma_{\lambda}, \Sigma_{\mu}] = i \varepsilon_{\lambda \mu \nu} \Sigma_{\nu}$, where $\varepsilon _{\lambda \mu \nu}$ is a completely antisymmetric tensor, with $\varepsilon_{xyz}=1$, and $\lambda, \mu, \nu = x,y,z$:
(1) ${\rm U(1)_{z}}$ generated by $K_z=\sum _j (-1)^{j+1}(S_j^z)^2$ and ${\rm SU(2)_{z,xy}}$ generated by
$\Sigma_x=\sum_j(-1)^{j+1}(S_j^xS_j^y+S_j^yS_j^x)/2$, $\Sigma_y=\sum_{j}S_j^z/2$ and  $\Sigma_z=K_{xy}/2$;
(2) ${\rm U(1)_{x}}$ generated by $K_x=\sum _j (-1)^{j+1}(S_j^x)^2$ and ${\rm SU(2)_{x,yz}}$ generated by
$\Sigma_x=\sum_j(-1)^{j+1}(S_j^yS_j^z+S_j^zS_j^y)/2$, $\Sigma_y=\sum_jS_j^x/2$ and $\Sigma_z=K_{yz}/2$;
(3) ${\rm U(1)_{y}}$ generated by $K_y=\sum _j (-1)^{j+1}(S_j^y)^2$ and ${\rm SU(2)_{y,zx}}$ generated by
$\Sigma_x=\sum_j(-1)^{j+1}(S_j^zS_j^x+S_j^xS_j^z)/2$, $\Sigma_y=\sum_jS_j^y/2$ and $\Sigma_z=K_{zx}/2$.

At the isotropic point $J_x=J_y=J_z$, a ${\rm SU(3)}$ symmetry group emerges, with its generators being
eight traceless operators
$J_1=1/2\sum_{j}S_j^x$, $J_2=1/2\sum_{j}S_j^{y}$, $J_3=1/2\sum_{j}S_j^z$,
$J_4=(-1)^{j+1}[1-3/2\sum_{j}(S_j^z)^2]$, $J_5=1/2\sum_{j}(-1)^{j+1}[{(S_j^x)}^2-{(S_j^y)}^2]$,
$J_6=1/2\sum_{j}(-1)^{j+1}(S_j^yS_j^z+S_j^zS_j^y)$, $J_7=1/2\sum_{j}(-1)^{j+1}(S_j^zS_j^x+S_j^xS_j^z)$ and
$J_8=1/2\sum_{j}(-1)^{j+1}(S_j^xS_j^y+S_j^yS_j^x)$.

This demonstrates that the spin-1 bilinear-biquadratic model~\cite{SMsutherland, SMTB, SMbarber, SMAffleck, SMChubukov, SMFath, SMKawashima,SMBatista,SMIvanov, SMBuchta, SMRizzi, SMLauchli, SMPorras, SMRomero, SMKluemper, SMSierra, SMThomale, SMRakov, SMdai0}, described by the Hamiltonian
\begin{equation}
	H= \sum _j (\cos{\theta}\;\vec{S}_j\cdot\vec{S}_{j+1}+\sin{\theta}\;(\vec{S}_j\cdot\vec{S}_{j+1})^2),\label{bb}
\end{equation}
possesses two extra $\rm{SU(3)}$ points: $\theta=\pm \pi/2$, in addition to the two $\rm{SU(3)}$ points at $\theta=\pi/4$ and $\theta=-3\pi/4$   (see also Refs. ~\cite{SMsu3affleck, SMxhchen}).   Actually, all the four points are exactly solvable.  In particular,  the Hamiltonian (\ref{bb}) with $\theta=\pi/4$ is the Uimin-Lai-Sutherland model~\cite{SMsutherland}, and the Hamiltonian (\ref{bb}) with $\theta=-\pi/2$ may be mapped into the nine-state Potts model~\cite{SMbarber}. In this sense, our investigation into an anisotropic extension of the spin-1 biquadratic model offers additional insights into the  spin-1 bilinear-biquadratic model (\ref{bb}), given that the spin-1 biquadratic model remains to be poorly understood up to the present.

\subsection{Duality transformations}\label{duality}
Quantum duality is local or nonlocal nontrivial unitary transformation $U$, which leaves the form of the local Hamiltonian density intact~\cite{SMfm, SMduality, SMgxyz}.

For simplicity, we choose $J_z$ as an energy scale by setting $J_z=1$.   That is, $J_x$ and $J_y$ are control parameters.
For a Hamiltonian $H(J_x,J_y)$, $H(J_x',J_y')$ is dual to $H(J_x,J_y)$, if there is a unitary transformation $U$ such that $H(J_x,J_y) = k'(J'_x,J'_y)UH(J'_x,J'_y) U^{\dagger}$, with $J'_x$ and $J'_y$ being some functions of $J_x$ and $J_y$, and $k'(J'_x,J'_y)$ being positive.
In particular, if $k'(J'_x,J'_y)=1$, then we refer to this particular case as a symmetric transformation of the Hamiltonian $H(J_x,J_y)$.

Here, we restrict ourselves to the region $J_x\geq0$ and $J_y\geq0$, due to the presence of a symmetric transformation: (1) $S^x_j\rightarrow(-1)^jS^x_j$, $S^y_j\rightarrow(-1)^jS^y_j$, $S^z_j\rightarrow S^z_j$, accompanied by $J_x\rightarrow J_x$, $J_y\rightarrow J_y$ and $J_z\rightarrow -J_z$; (2) $S^x_j\rightarrow S^x_j$, $S^y_j\rightarrow (-1)^jS^y_j$, $S^z_j\rightarrow (-1)^j S^z_j$, accompanied by $J_x\rightarrow -J_x$, $J_y\rightarrow J_y$ and $J_z\rightarrow J_z$;
(3) $S^x_j\rightarrow (-1)^j S^x_j$, $S^y_j\rightarrow S^y_j$, $S^z_j\rightarrow (-1)^j S^z_j$, accompanied by $J_x\rightarrow J_x$, $J_y\rightarrow -J_y$ and $J_z\rightarrow J_z$.
In addition, we have one extra symmetric transformation and two duality transformations:

(0) The Hamiltonian $H(J_x,J_y)$ is symmetric under a local unitary transformation $U_0$:
$S_j^x\rightarrow S_j^y$,
$S_j^y\rightarrow S_j^x$ and $S_j^z\rightarrow-S_j^z$,
accompanied by $J_x \leftrightarrow J_y$.

(1) Under a local unitary transformation $U_1$:
$S_{j}^x\rightarrow S_{j}^z$,
$S_{j}^y\rightarrow-S_{j}^y$, and $S_{j}^z\rightarrow S_{j}^x$, we have
$H(J_x, J_y)=k'(J'_x, J'_y) U_1 H(J'_x, J'_y) U^\dagger_1$,
with $J_x=1/J'_x$, $J_y=J'_y/J'_x$, and $k'(J'_x,J'_y) =1/J'^{2}_x$.
The Hamiltonian on the line $J_x=1$ is self-dual.

(2) Under a local unitary transformation $U_2$:
$S_{j}^x\rightarrow-S_{j}^x$,
$S_{j}^y\rightarrow S_{j}^z$, and $S_{j}^z\rightarrow S_{j}^y$,
we have $H(J_x, J_y)=k'(J'_x, J'_y) U_2 H(J'_x, J'_y) U^\dagger_2$,
with $J_x=J'_x/J'_y$, $J_y=1/J'_y$, and $k'(J'_x,J'_y) =1/J'^{2}_y$.
The Hamiltonian on the line $J_y=1$ is self-dual.\\

\begin{figure}
	\includegraphics[angle=0,totalheight=4cm]{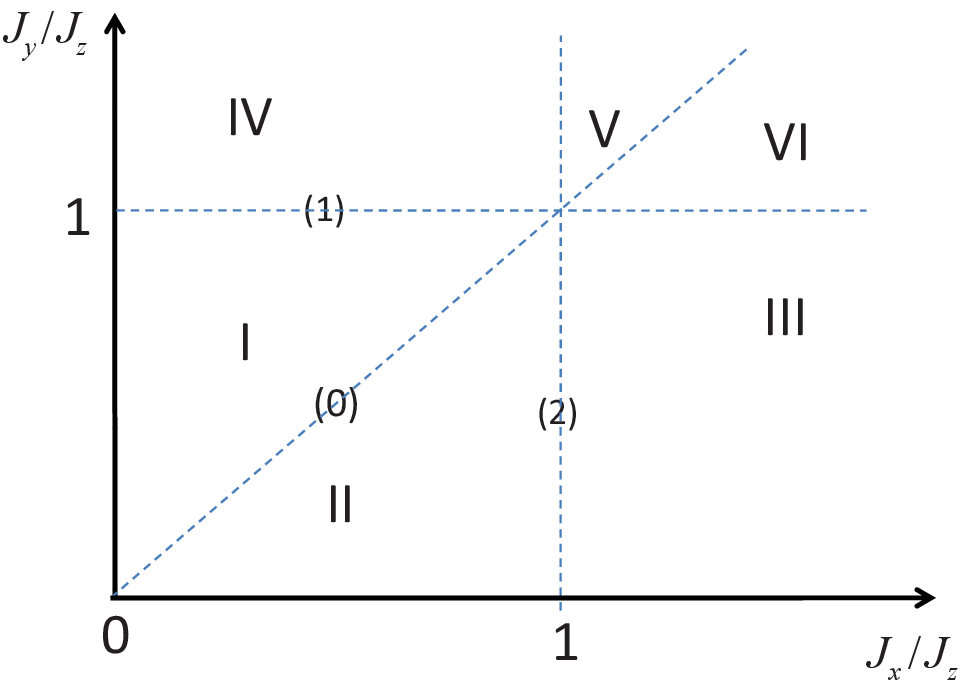}
	\caption{ Six dual regimes generated from one symmetric transformation and two duality transformations for an anisotropic extension of the spin-$1$ biquadratic model in the region: $J_x/J_z \geq 0$ and $J_y/J_z \geq 0$. }\label{sixregimes}
\end{figure}

As a consequence, the entire parameter region, as shown in Fig.~\ref{sixregimes}, is divided, via the three characteristic lines: $J_x=J_y$, $J_y=J_z$ and $J_z=J_x$, into six different regimes, which are dual to each other.

\subsection{Highly degenerate ground states on the two characteristic lines}\label{twolines}

We present the exact ground states of the Hamiltonian (1) on the two characteristic lines:  $J_x=0$, with $J_y/J_z > 0$ and  $J_y=J_z$, with $J_x/J_z<1$.

On the characteristic line $J_x=0$, with $J_y/J_z > 0$, there exists a two-parameter family of factorized ground states for fixed $J_y/J_z$, with the ground state energy per site $e$ being equal to zero: $|\Psi_f\rangle=\bigotimes_m |v_1v_2\rangle _m$, where $|v_1v_2\rangle _m = |v_1\rangle_{2m-1} |v_2\rangle_{2m}$, with $|v_1\rangle_{2m-1}$ and $|v_2\rangle_{2m}$ being a vector in a local spin space at lattice sites $2m-1$ and $2m$, respectively.
The explicit expressions for $|v_1\rangle_{2m-1}$ and $|v_2\rangle_{2m}$ have been presented in Eq.~(3).

It is found that $|\Psi_f\rangle$ becomes the translation-invariant factorized state $|{\rm TIGS}\rangle$, if $p=\sqrt{J_y/(J_y+J_z)}$ and $\delta=0$ in Eq.~(2).
A sequence of factorized ground states $|\Phi_f(\alpha)\rangle$
are generated from the action of the symmetry group element $V$ on $|{\rm TIGS}\rangle$: $|\Phi_f(\alpha)\rangle\propto V|{\rm TIGS}\rangle$, where $V=\exp{(\alpha K_{yz})}$, with $\alpha=\zeta+i \varepsilon$. Here, $\zeta$ and $\varepsilon$ are two real free parameters.
That is, we have
\begin{equation}
	|\Phi_f(\alpha)\rangle=\bigotimes_m|\upsilon_1 \upsilon_2 \rangle_m ,
	\label{phifv}
\end{equation}
with $|\upsilon_1\upsilon_2\rangle _m = |\upsilon_1\rangle_{2m-1} |\upsilon_2\rangle _{2m}$.
Specifically, if $\alpha$ is pure imaginary: $\alpha=i\varepsilon$,  then the operator $V$ is unitary, thus yielding a factorized ground state $|\Phi_f(\alpha)\rangle$,  equivalent to a factorized ground state in Eq.~(2), with $p=\sqrt{J_y/(J_y+J_z)}$ and $\delta=-2\varepsilon$.  In other words, $|\upsilon_1\rangle$ and $|\upsilon_2\rangle$ are in the form
\begin{eqnarray*}
	|\upsilon_1\rangle_{2m-1}\;=\;&\frac{1}{\sqrt{J_y+J_z}} (\;\sqrt{J_y}|0_y\rangle_{2m-1}\;+\;\exp{(-2i\varepsilon)}\sqrt{J_z}|0_z\rangle_{2m-1}),\\
	|\upsilon_2\rangle_{2m}=&\frac{1}{\sqrt{J_y+J_z}} (\;\sqrt{J_y}|0_y\rangle_{2m}+\exp{(2i\varepsilon)}\sqrt{J_z}|0_z\rangle_{2m}).
\end{eqnarray*}
Generically, if $\alpha=\zeta+i\varepsilon$, then $|\Phi_f(\alpha)\rangle$ is a factorized ground state in Eq.~(2), with $p=\sqrt{J_y/[J_y+\exp(-4\zeta)J_z]}$ and $\delta=-2\varepsilon$.  In other words, $|\upsilon_1\rangle$ and $|\upsilon_2\rangle$ are in the form
\begin{eqnarray*}
	|\upsilon_1\rangle_{2m-1}=&\frac{1}{\sqrt{N_1}} (\sqrt{J_y}|0_y\rangle_{2m-1}+\exp{(-2\zeta)}\exp{(-2i\varepsilon)}\sqrt{J_z}|0_z\rangle_{2m-1}),\\
	|\upsilon_2\rangle_{2m}=&\frac{1}{\sqrt{N_2}} (\sqrt{J_y}|0_y\rangle_{2m}+\exp{(2\zeta)}\exp{(2i\varepsilon)}\sqrt{J_z}|0_z\rangle_{2m}).
\end{eqnarray*}
Here, $N_1=J_y+\exp(-4\zeta)J_z$ and $N_2=J_y+\exp(4\zeta)J_z$.
In particular, when $\zeta\rightarrow\infty$ and $\varepsilon=0$, we have $|\upsilon_1\rangle_{2m-1}=|0_y\rangle_{2m-1}$ and $|\upsilon_2\rangle_{2m}=|0_z\rangle_{2m}$.  We stress that this operation is not invertible, indicating that the limiting procedure is singular.

It is straightforward to evaluate the ground state fidelity per site for two factorized ground states.
This implies that any two factorized ground states are orthogonal to each other in the thermodynamic limit. Physically, this suggests that the symmetry group ${\rm U}(1)$, generated by $K_{yz}$,
is spontaneously broken, in both unitary and non-unitary realizations.

On the characteristic line: $J_y=J_z$ with $J_x/J_z<1$, the symmetry group ${\rm SU(2)_{x,yz}}$, generated by
$\Sigma_x=\sum_j(-1)^{j+1}(S_j^yS_j^z+S_j^zS_j^y)/2$, $\Sigma_y=\sum_jS_j^x/2$, and $\Sigma_z=K_{yz}/2$, emerges, together with the symmetry group ${\rm U(1)_{x}}$, generated by $K_{x}=\sum_j (-1)^{j+1}(S_j^x)^2$. Since the one-site translation-invariant factorized ground state $|{\rm TIGS}\rangle$ in Eq.~(2) becomes $|{\rm TIGS}\rangle=\bigotimes_j|1_x\rangle_j$, with $p=q=s=t=\sqrt{2}/2$, on the characteristic line $J_y=J_z$.
Here, $|1_x\rangle_j$ is a basis state, with an eigenvalue being one, for the spin-$1$ operator $S^x_j$.
Note that $|{\rm TIGS}\rangle=\bigotimes_j|1_x\rangle_j$ is the highest weight state for the symmetry group $\rm{SU(2)}$, generated by $\Omega_x=\Sigma_z$, $\Omega_y=\Sigma_x$, and $\Omega_z=\Sigma_y$.
Accordingly, the raising operator $\Omega_+=\sum_j{\Omega_{+,j}}$ and the lowering operator $\Omega_-=\sum_j{\Omega_{-,j}}$ are defined as $\Omega_{\pm,j}=(\Omega_{x,j}\pm i\Omega_{y,j})/\sqrt{2}$, respectively: $[\Omega_z,\Omega_+]=\Omega_+$, $[\Omega_+,\Omega_-]=\Omega_z$ and $[\Omega_-,\Omega_z]=\Omega_-$.
Therefore,  a sequence of the degenerate ground states $|\Theta(L,N)\rangle$ on the characteristic line $J_y=J_z$ are generated from the repeated action of the lowering operator $\Omega_-$ on the highest weight state $|{\rm TIGS}\rangle=\bigotimes_j|1_x\rangle_j$.
In Ref. \cite{SMFDGM}, a systematic investigation has been performed for highly degenerate and highly entangled ground states on the characteristic line, which arises from SSB with type-B GMs.
It is found that the entanglement entropy $S(n)$ for the degenerate ground states scales logarithmically with the block size $n$ in the thermodynamic limit, with the prefactor being half the number of type-B GMs $N_B$. The latter in turn is identical to the fractal dimension~\cite{SMdoyon}.
According to the counting rule, SSB from ${\rm SU(2)}\times{\rm  U(1)}$ to ${\rm U(1)}\times{\rm U(1)}$ leads to one type-B GM~\cite{SMwatanabe}: $N_B=1$. Therefore, the fractal dimension $d_f$ is identified to be $d_f=1$.

In addition, there are two other factorized ground states (i) $|\Phi_f^{yz}\rangle=\bigotimes_m|0_y0_z\rangle_m$ and (ii) $|\Gamma_f\rangle=\bigotimes_m (|0_y\rangle_{2m-1}-i|0_z\rangle_{2m-1}) (|0_y\rangle_{2m}+i|0_z\rangle_{2m})$, which act as the highest weight states for the symmetry group $\rm{SU(2)}$, generated by $\Sigma_x$, $\Sigma_y$, and $\Sigma_z$,  and $\Xi_x=\Sigma_y$, $\Xi_y=\Sigma_z$, and $\Xi_z=\Sigma_x$, respectively.
Accordingly, other two sequences of the degenerate ground states may be generated in the same way. However, the three sequences are unitarily equivalent to each other.
As a consequence, the entanglement entropy $S(n)$ for the three sequences of the highly degenerate ground states must be identical.

At the isotropic point ($J_x=J_y=J_z$), SSB from $\rm{SU(3)}$ to ${\rm U(1)}\times{\rm U(1)}$ occurs~\cite{SMstaggeredsu3}, with six of the eight generators for the (staggered) symmetry group $\rm{SU}(3)$ being spontaneously broken. This leads to two type-B GMs, with two broken generators being redundant, as required to keep consistency with the counting rule~\cite{SMwatanabe}.
Accordingly, the entanglement entropy $S(n)$ for the degenerate ground states scales logarithmically with the block size $n$ in the thermodynamic limit, with the prefactor being half the number of type-B GMs $N_B$. Therefore, the fractal dimension $d_f$ is identified to be $d_f=2$.

\subsection{Highly degenerate ground states in the coexisting fractal phases} \label{u1ssb}

In the ${\rm CF}_x$ phase, away from the characteristic line $J_y=J_z$, the Hamiltonian possesses the symmetry group $\rm{U(1)}\times\rm{U(1)}$.  That is,  the symmetry group ${\rm SU(2)} \times {\rm U(1)}$
is explicitly broken to  $\rm{U(1)}\times\rm{U(1)}$. As a consequence, a gapped GM emerges. However, as already indicated when we discussed the exact factorized ground states on the characteristic line $J_x=0$, an exotic type of SSB from ${\rm U(1)} \times {\rm U(1)}$ to ${\rm U(1)}$ arises, leading to highly degenerate (factorized) ground states in the thermodynamic limit~\cite{SMU1SSB}.

\begin{figure}
   \includegraphics[width=0.35\textwidth]{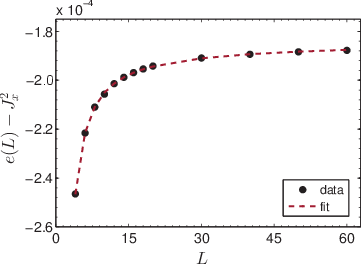}
	\caption{(color online) The finite-size corrections to the ground state energy per lattice site $e(L)$, for the chosen point $(J_x/J_z, J_y/J_z)=(0.4, 0.94)$ in the ${\rm CF}_x$ phase. Here, the finite-size DMRG algorithm is exploited to simulate the model~(1) under PBCs, with $L$ ranging from $4$ to $60$.}
	\label{Efsc}
\end{figure}

A finite-size scaling analysis for the ground state energy per site $e(L)$ is performed in the $\rm{CF_x}$ phase, with $L$ being the size. As argued in Ref.~\cite{SMU1SSB},
the finite-size corrections to the ground state energy per site $e(L)$ take the form in Eq.~(4). 
Indeed, the finite-size corrections to the ground state energy per site may be justified from the fact that there are a sequence of $q$ ${\rm H}$-orthogonal states, with $q=L+1$~\cite{SMU1SSB}~\cite{SMszb}.
Here, $-A\exp{(\eta/ L)}/L$ arises from the emergent permutation symmetry in the ground state subspace,
which in turn is relevant to a gapped GM when the symmetry group ${\rm SU}(2)\times{\rm  U}(1)$ on the characteristic line $J_y=J_z$ is explicitly broken to ${\rm  U(1)}\times{\rm  U(1)}$, and $-B\exp(-\kappa L)$ represents a contribution from an alternative SSB pattern for ${\rm  U(1)}$ generated by $K_{yz}$.
This prediction is confirmed numerically by means of the finite-size DMRG simulations~\cite{SMU1SSB}.
The typical values of $A$, $B$ and $\kappa$ are at the order of magnitude $10^{-4}$, whereas $\eta$ is around 1.
Here, we take $J_x/J_z=0.4$ and $J_y/J_z=0.94$ as an example, as shown in Fig.~\ref{Efsc}. The best fit for $e(L)$ results in  $A=1.728\times 10^{-4}$, $B=1.853\times 10^{-4}$, $\eta=1.4$ and $\kappa=0.655\times 10^{-4}$.
In fact, the finite-size corrections to the ground state energy per site $e(L)$ are always negative, implying that the ground state energy per site $e(L)$ in the $\rm{CF_{x}}$ phase approaches $J_x^2$ from below, as $L$ tends to infinity. Note that the entanglement entropy for a ground state wave function is non-zero, when the size $L$ is finite. We anticipate that it survives even in the thermodynamic limit, because  the ground state wave function remains to be a linear combination of the symmetry-broken states arising from SSB for the symmetry group ${\rm  U}(1)$ generated by $K_{yz}$.

\begin{figure}
	\vspace{10pt}
	\includegraphics[angle=0,width=0.35\textwidth]{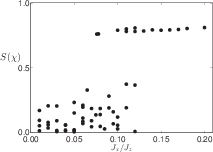}\\
	\caption{ The entanglement entropy $S(\chi)$ as a function of $J_x/J_z$ for fixed $J_y/J_z=0.7$, as a result of the iTEBD simulations, with the bond dimension $\chi=48$.  Here, different initial states have been chosen randomly for each value of $J_x/J_z$. The entanglement entropy $S(\chi)$ exhibits different behaviors for the two distinct phases: it takes a definitive value in the $\rm{LL_{yz}}$ phase, and a random value in the $\rm{CF_x}$ phase, respectively.}\label{entropy2}
\end{figure}

\begin{figure}
	\vspace{10pt}
	\includegraphics[angle=0,width=0.35\textwidth]{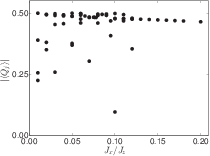}
	\caption{ For fixed $J_y/J_z=0.7$, the amplitude $|\langle Q_j \rangle|$ exhibits distinct behaviors to distinguish the $\rm{LL_{yz}}$ phase from the $\rm{CF_x}$ phase. Here, we have chosen the bond dimension $\chi$ to be  $\chi=48$. }\label{order2}
\end{figure}

This explains why the ground state energy per site from the iTEBD simulations for an accessible value of the bond dimension $\chi$ is always lower than $J_x^2$ in the $\rm{CF_{x}}$ phase, with the magnitude of the deviation being at the order of magnitude $10^{-4}$, which in turn is at the same order of magnitude as $A$ and $B$ in Eq.~(4). This is due to the fact that a finite length scale emerges due to the finiteness of the bond dimension $\chi$. As a result,  the entanglement entropy $S(\chi)$ in the $\rm{CF_{x}}$ phase does not vanish for a ground state wave function generated from the iTEBD simulations.
We plot the entanglement entropy $S(\chi)$ as a function of $J_x/J_z$ for fixed $J_y/J_z=0.7$ in Fig.~\ref{entropy2}, with the bond dimension $\chi=48$.
Here, different initial states have been chosen randomly for each value of $J_x/J_z$.
A remarkable fact is that the entanglement entropy $S(\chi)$ exhibits different behaviors for the two distinct phases: it takes a definitive value in the $\rm{LL_{yz}}$ phase, and a random value in the $\rm{CF_x}$ phase, respectively.
In fact, this observation may be used to signal QPTs from the $\rm{LL_{yz}}$ phases to the $\rm{CF_x}$ phases.  Randomness also occurs in the amplitude of the (pseudo) local order parameter $\langle Q_j\rangle$.
As shown in Fig.~\ref{order2}, for fixed $J_y/J_z=0.7$, the amplitude $|\langle Q_j\rangle|$ exhibits distinct behaviors for the two phases:
in the $\rm{LL_{yz}}$ phase, $|\langle Q_j\rangle|$ varies smoothly with $J_x/J_z$, whereas in the $\rm{CF_x}$ phase,
$|\langle Q_j\rangle|$ varies randomly with $J_x/J_z$. In practice, this distinction makes it possible to distinguish the $\rm{CF_x}$ phase from the $\rm{LL_{yz}}$ phase.

\subsection{The non-local order parameters for the SPt phases}\label{nonlocalspt}

Powerful tensor network algorithms~\cite{SMvidal,SMidmrg} may be exploited to efficiently simulate a quantum many-body system in one spatial dimension.
The algorithms generate ground state wave functions in the iMPS representation on an infinite-size chain, and provides an efficient means to evaluate various physical observables.

\begin{figure}
	\includegraphics[angle=0,width=0.45\textwidth]{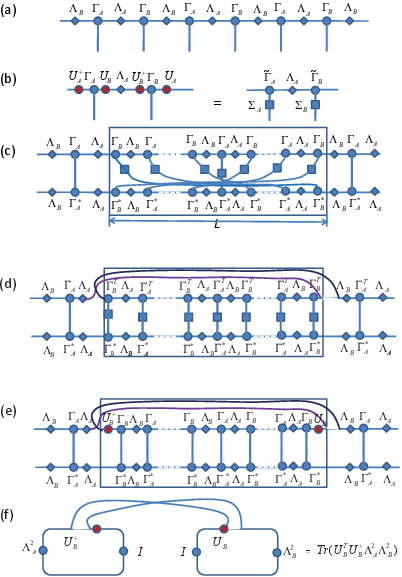}\\
	\caption{ Diagrammatic derivation of the non-local order parameters $\langle T_{\mu} \rangle$ $(\mu =x,y,z)$.
		(a) Two-site translation-invariant iMPS representation consisting of $\Gamma_A$, $\Gamma_B$, $\Lambda_A$ and $\Lambda_B$;
		(b) A pictorial representation of the unitary transformations $U_A$ and $U_B$ induced from a symmetry group $G$ to connect an original wave function to the transformed wave function under the symmetric transformation $\Sigma_A \in G$ and $\Sigma_B \in G$, together with the site-centered inversion symmetry.  Here,  $\tilde{\Gamma}_A$ and $\tilde{\Gamma}_B$ represent $\Gamma_A$ and $\Gamma_B$ under the site-centered inversion symmetric transformation, and $\Sigma_A$ and $\Sigma_B$ represent a $\pi$-rotation around the $\mu$-axis in the spin space, with $\mu =x,y,z$.
		(c) The overlap of a wave function with a partially inverted wave function for an infinite-size chain, in which a segment of $2l+1$ sites have been inverted.
		(d) The overlap is untwisted by reversing the segment consisting of $2l+1$ sites.
		(e) Recover $\Gamma_A$ and $\Gamma_B$ by inserting the unitary transformations $U_A$ and $U_B$.
		(f) For a large $l$, the non-local order parameter $\langle T_{\mu} \rangle$ is simplified by keeping only the dominant eigenvector of the transfer matrix, with the largest eigenvalue.    As a result, we have $\langle T_{\mu} \rangle ={\rm Tr}(U_B^TU^\dagger_B\Lambda_A^2\Lambda_B^2)/{\rm Tr}(\Lambda_A^2\Lambda_B^2)$.
	}\label{nonlocald}
\end{figure}

A conventional choice is the two-site translation-invariant iMPS representation for a ground state wave function when the model Hamiltonian is either one-site or two-site translation-invariant, as shown in Fig.~\ref{nonlocald} (a).
The bond dimension $\chi$ in the iMPS representation imposes an upper bound on the bipartite entanglement present in a given ground state wave function: $\log_2 \chi$.
Hence, such a representation is efficient for ground states in a gapped phase, as long as the bond dimension $\chi$ is large enough.
Meanwhile, a finite-entanglement scaling analysis provides a practical means to characterize a gapless phase at criticality.

As an illustration, we discuss how to evaluate a non-local order parameter for a SPt phase.  Following Fuji, Pollmann, and Oshikawa~\cite{SMpollmann2}, the combined operation of the site-centered inversion symmetry with a $\pi$-rotation in the spin space is introduced to characterize a SPt phase, which is a symmetric phase connected adiabatically to a product state.
As argued~\cite{SMpollmann2}, such a SPt phase is different from a SPT phase~\cite{SMwenxg,SMpollmann}.

For our purpose, it is convenient to introduce three non-local order parameters $\langle T_{\mu} \rangle$, defined by the combined operation of the site-centered inversion symmetry $I$ with a $\pi$-rotation about the $\mu$ axis, $R_\mu = \exp{(i\pi S^\mu)}$, with $\mu = x, y$, and $z$~\cite{SMpollmann2}. That is, $T_\mu=I \cdot R_\mu$.  If the combined symmetry is retained, $\langle T_\mu \rangle$ take either $1$ or $-1$.   The iMPS representation offers us a diagrammatic derivation of the non-local order parameters $\langle T_\mu \rangle$, as shown in Fig.~\ref{nonlocald}, with $\Sigma$ being a $\pi$-rotation $R_\mu$ around the $\mu$ axis in the spin space.
For an anisotropic extension of the spin-1 biquadratic model,  there are three distinct SPt phases, which are characterized by a set of the non-local order parameters $(\langle T_x\rangle,\langle T_y\rangle,\langle T_z\rangle)$.

\begin{figure}
	\includegraphics[angle=0,width=0.35\textwidth]{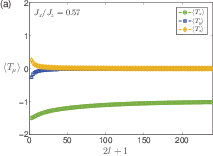}\\
	\vspace{0.5cm}
	\includegraphics[angle=0,width=0.35\textwidth]{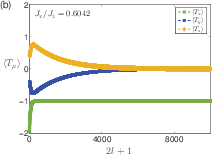}\\
	\vspace{0.5cm}
	\includegraphics[angle=0,width=0.35\textwidth]{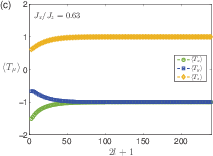}
	\caption{ Saturation of the non-local order parameters $\langle T_\mu \rangle$ ($\mu = x,y,z$) with the block size $2l+1$	for two typical points: (a) $(J_x/J_z=0.57,J_y/J_z=0.7)$ and (b) $(J_x/J_z=0.6042,J_y/J_z=0.7)$ in the $\rm{LL_{yz}}$ phase and one typical point: (c) $(J_x/J_z=0.63,J_y/J_z=0.7)$ in the $\rm{SPt_{z}}$ phase. Here, the bond dimension $\chi$ is chosen to be 60. }\label{nonlocalorder}
\end{figure}

For fixed $J_y/J_z=0.7$, we choose two typical values of $J_x/J_z$ in the $\rm{LL_{yz}}$  phase: $J_x/J_z=0.57$ and $0.6042$  and one typical value of $J_x/J_z$  in the $\rm{SPt_{z}}$ phase: $J_x/J_z=0.63$.  In Fig.~\ref{nonlocalorder} (a), (b) and (c), we plot the non-local order parameters $\langle T_\mu \rangle $ as a function of the block size $2l+1$, respectively, 
with the bond dimension $\chi=60$.  This suggests that a large block size is needed to ensure that the non-local order parameters $\langle T_\mu \rangle$ are saturated, if a phase transition point is approached.

\subsection{The LL phases and the pseudo local order parameters}\label{pseudoorderparameterLL}

The KT transitions describe an instability of the LLs, due to marginal perturbations, towards a gapful phase, with or without $Z_2$ SSB order.  Normally, it is not an easy task to determine whether or not the KT transitions occur in a specific quantum many-body system, due to the fact that the KT transitions exhibit essential singularities and no local order parameter exists, since no symmetry is spontaneously broken in a LL phase.
As a consequence, widely used finite-size scaling techniques fail in characterizing the KT transitions.   However, the situation is quite subtle
when one performs numerical simulations in terms of the iTEBD algorithms. In particular,  the algorithms yield an infinite number of degenerate ground states in a LL phase, thus resulting in a pseudo local order parameter, if the translational invariance is retained.

To understand how such a pseudo local order parameter arises in the iMPS representation of ground state wave functions for quantum many-body systems in one spatial dimension, we recall the notion of continuous SSB in the conventional Landau-Ginzburg-Wilson paradigm.
Although the system Hamiltonian possesses a certain continuous symmetry, the ground states for the system do not satisfy the symmetry, which leads to the occurrence of SSB.
Such a symmetry breakdown originates from a random perturbation, and results in an infinite number of degenerate ground states.
In principle, SSB only occurs for quantum many-body systems in higher than one spatial dimensions, as follows from the Mermin-Wagner-Coleman theorem.
However, in the iMPS representation for the iTEBD algorithm, the finiteness of the bond dimension $\chi$ results in a finite gap, which vanishes when $\chi\rightarrow \infty$.
As a consequence, pseudo continuous SSB occurs in quantum many-body systems in one spatial dimension, resulting in a pseudo local order parameter~\cite{SMdai0, SMwang}.
This offers us a novel characterization of the KT transitions in terms of pseudo continuous SSB, which in turn makes it possible to introduce a pseudo local order parameter, just as in the conventional Landau-Ginzburg-Wilson paradigm.  In this scenario, a LL phase may be characterized as a limiting case of the continuous SSB in the conventional Landau-Ginzburg-Wilson paradigm.

In fact, a pseudo local order parameter tends to vanish when the bond dimension $\chi$ increases, as required to keep consistency with the Mermin-Wagner-Coleman theorem.
In addition, it is argued in Refs.~\cite{SMdai0, SMwang} that, for any finite choice of the bond dimension $\chi$, pseudo continuous SSB is reflected as
a catastrophe point in the ground state fidelity per site~\cite{SMzhou} for quantum many-body systems in one spatial dimension,
undergoing the KT transitions.
In fact, if $\chi$ tends to infinity, then such a catastrophe point turns into an essential singularity.

In practice, pseudo SSB may be recognized from the fact that, for any finite value of the bond dimension $\chi$, a ground state wave function depends on an initial state chosen randomly, and shares all the features of the continuous SSB:
(1) a system has stable and degenerate ground states, each of which breaks the symmetry
of the system; 
(2) the symmetry breakdown results from random perturbations;
(3) such (pseudo) symmetry-breaking order may be quantified by introducing a (pseudo) local order
parameter, which may be read off from the reduced density matrix on a local area~\cite{SMzhou0}.

For the model under investigation,  it possesses the symmetry group $\rm{U(1)}\times\rm{U(1)}$  in the entire parameter space.
In the LL phases, it is found that, pseudo SSB occurs for one of the $\rm{U(1)}$ symmetry groups, as a result of the finiteness of the bond dimension $\chi$ in the iMPS representation.
For the ${\rm LL}_{yz}$ phase, this means that a pseudo local order parameter $\langle Q_j \rangle$ emerges: $\langle Q_j\rangle$ takes different values for different degenerate ground states, generated from a randomly chosen initial state, when the iTEBD algorithm is implemented.  This is seen from the fact that $\langle WQ_jW^{\dag}\rangle=\exp (i\; 2\varepsilon)\langle Q_j\rangle$ for $W=\exp{(i\varepsilon K_{yz})}$.
Note that, the real and imaginary parts of the  pseudo local order parameter $\langle Q_j\rangle$ are $ \langle (-1)^{j+1}(S^y_jS^z_j+S^z_jS^y_j)/2 \rangle$ and $ \langle S^x_j/2 \rangle$, respectively.  This suggests that pseudo SSB for the unitary realization of the symmetry group ${\rm U}(1)$, generated by $K_{yz}$, occurs in the $\rm{LL_{yz}}$ phase.

\begin{figure}
	\vspace{5mm}
	\includegraphics[angle=0,width=0.4\textwidth]{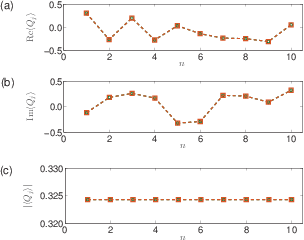}
	\caption{ The pseudo order parameter $\langle Q_{j}\rangle$ is evaluated from ground state wave functions, which are generated from ten randomly chosen trial initial states, labeled by $n$, in terms of the iTEBD simulations. Here, the simulations are performed for the Hamiltonian (1) with $J_x/J_z=0.57$ and $J_y/J_z=0.7$, which is located in the ${\rm LL}_{yz}$ phase, and the bond dimension $\chi$ is chosen to be 48.
	(a) The real parts {\rm Re}$\langle Q_{2m-1}\rangle$ (circles) at odd sites and {\rm Re}$\langle Q_{2m}\rangle$ (squares) at even sites vary as $n$  changes. (b) The imaginary parts {\rm Im}$\langle Q_{2m-1}\rangle$ (circles) at odd sites and {\rm Im}$\langle Q_{2m}\rangle$ (squares) at even sites vary as $n$ changes. (c) The amplitudes $|\langle Q_{2m-1}\rangle|$ (circles) at odd sites and $|\langle Q_{2m}\rangle|$ (squares)  at even sites vary as $n$ changes.
	It is found that {\rm Re}$\langle Q_j\rangle$ and {\rm Im}$\langle Q_j\rangle$ depend on $n$, but $|\langle Q_j\rangle|$ does not, within a reasonable error. As a result, pseudo SSB occurs in the ${\rm LL}_{yz}$ phase.
	}\label{porderLL}
\end{figure}

In Fig.~\ref{porderLL}, we plot the pseudo local order parameter $\langle Q_j \rangle$
for $J_x/J_z=0.57$ and $J_y/J_z=0.7$, which is located in the ${\rm LL}_{yz}$ phase, with the bond dimension $\chi=48$, for degenerate ground states generated from ten different initial states.
It is found that the real and imaginary parts, {\rm Re}$\langle Q_j\rangle$ and {\rm Im}$\langle Q_j\rangle$, are different, but the amplitude $|\langle Q_j\rangle|$ remains the same for fixed $J_x/J_z$ and $J_y/J_z$.
From the invariance of the iMPS representation under the two-site translation operation, the real and imaginary parts of  $\langle Q_j \rangle$ are $ \langle (-1)^{j+1}(S^y_jS^z_j+S^z_jS^y_j)/2 \rangle$ and $ \langle S^x_j/2 \rangle$, respectively. That is, for the real part, the local operator is alternating between odd and even sites, whereas for the imaginary part, the local operator is uniform. 
This implies that an infinite number of degenerate ground states are generated from randomly chosen initial states, as a result of pseudo SSB, due to the finiteness of the bond dimension $\chi$.

\subsection{Central charge $c$ in the ${\rm LL}_{yz}$ phase: a finite-size approach}\label{centralcf}

\begin{figure}[htbp]
	\includegraphics[angle=0,width=0.45\textwidth]{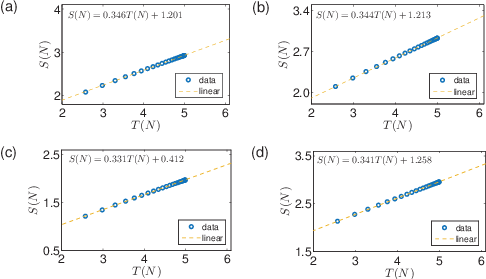}
	\vspace{4pt}
	\includegraphics[angle=0,width=0.24\textwidth]{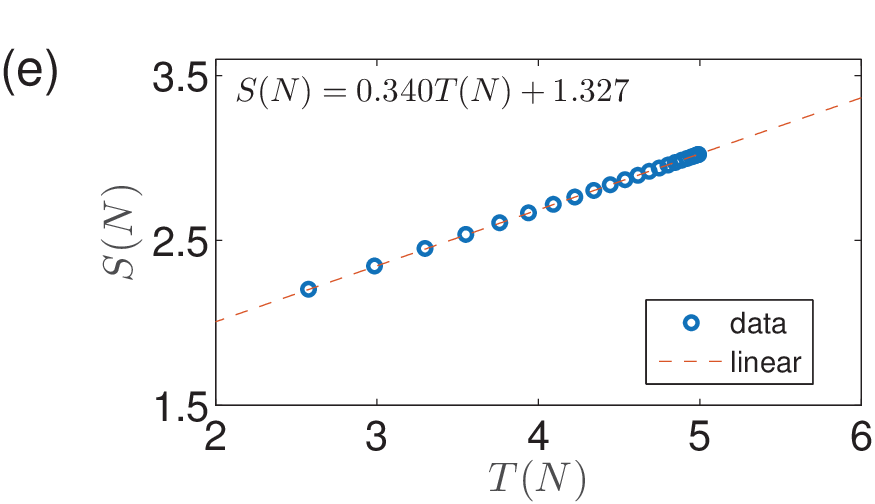}
	\caption{ The entanglement entropy $S(N)$ versus $T(N) \equiv \log_2{[L/\pi\sin(\pi N/L)]}$ for five chosen points in the ${\rm LL_{yz}}$ phase: (a) $J_x/J_z=0.15$ and  $J_y/J_z=0.5$; (b) $J_x/J_z=0.3$ and $J_y/J_z=0.6$; (c) $J_x/J_z=0.2$ and $J_y/J_z=0.7$; (d)  $J_x/J_z=0.5$ and  $J_y/J_z=0.7$; and (e) $J_x/J_z=0.7$ and $J_y/J_z=0.8$.  Here, we have chosen the size $L=100$ and the bond dimension $\chi=40$.   }\label{Sfinite}
\end{figure}

If a quantum many-body system under the periodic boundary conditions, with the size being $L$, is partitioned into a subsystem $N$ and its environment $L-N$,
a prediction from conformal field theory~\cite{SMcft} implies that the entanglement entropy $S(N)$ scales as
\begin{equation}
	S(N)=\frac{c}{3}T(N)+S_0,
\end{equation}
with $T(N)=\log_2{[L/\pi\sin(\pi N/L)]}$. Here, $S_0$ is a (model-dependent) additive constant.
Obviously, we have $S(N)=S(L-N)$.
A numerical simulation is performed in terms of the variational MPS algorithm~\cite{SMfrank}, which yields the ground state wave functions for a finite-size quantum many-body system under the periodic boundary conditions.
In Fig.~\ref{Sfinite}, a scaling analysis between the entanglement entropy $S(N)$ and $T(N)$ is performed for five choices of the coupling parameters $J_x/J_z$ and $J_y/J_z$: (a) $J_x/J_z=0.15$ and  $J_y/J_z=0.5$; (b) $J_x/J_z=0.3$ and $J_y/J_z=0.6$; (c) $J_x/J_z=0.2$ and $J_y/J_z=0.7$; (d)  $J_x/J_z=0.5$ and  $J_y/J_z=0.7$; and (e) $J_x/J_z=0.7$ and $J_y/J_z=0.8$, in the ${\rm LL_{yz}}$ phase, with the size $L=100$ and the bond dimension $\chi=40$.
In Table~\ref{tab2}, central charge $c$ is extracted from the five chosen points in the ${\rm LL_{yz}}$ phase.
As is seen, central charge $c$ is close to the exact value $c=1$, with a relative error being less than $4\%$.

\begin{table}[htbp]
	\centering
	\caption{Central charge $c$, extracted from the finite-size scaling, for five chosen points in the $\rm{LL_{yz}}$ phase.}
	\vspace{3mm}
	\label{tab2}
	\begin{tabular}{|c|c|c|c|c|c|}
		\hline
		($J_x/J_z$,$J_y/J_z$)&(0.15,0.5)&(0.3,0.6)&(0.2,0.7)&(0.5,0.7)&(0.7,0.8)\\
		\hline
		$c$&1.038&1.032&0.993&1.023&1.020 \\
		\hline
	\end{tabular}
\end{table}

\subsection{ Acknowledgement} 

 We acknowledge enlightening discussions with Murray Batchelor, Sam Young Cho, John Fjaerestad, Javier Rodriguez Laguna, Ian McCulloch, and German Sierra.

\end{document}